\documentclass[conference]{IEEEtran}
\usepackage[utf8]{inputenc}
\usepackage{cite}
\usepackage[english]{babel}
\usepackage{epsfig}
\usepackage{algorithm}
\usepackage{algorithmic}
\usepackage{amsmath,mathtools}
\usepackage{amsthm}
\usepackage{amssymb,bm}
\usepackage{dsfont}
\usepackage{subcaption}
\usepackage{balance}
\usepackage{pdflscape}
\usepackage{placeins}
\usepackage{graphicx}
\usepackage{cases}
\usepackage[figurename=Fig.]{caption}
\usepackage{amsfonts}
\usepackage{multirow}
\usepackage{multicol}
\usepackage{mathrsfs}
\usepackage{color}
\usepackage{soul,comment}
\usepackage{flushend}
\usepackage{subcaption}
\usepackage[table,xcdraw]{xcolor}
\usepackage{colortbl}
\usepackage{booktabs}

\usepackage{balance}
\usepackage{url}
\usepackage{epstopdf}
\usepackage{longtable}
\allowdisplaybreaks
\usepackage{array}
\usepackage{upgreek}
\newcolumntype{P}[1]{>{\centering\arraybackslash}p{#1}}

\usepackage{xcolor}
\usepackage{algorithm}
\usepackage{float}
\usepackage{graphicx}
\usepackage[normalem]{ulem}

\usepackage[font=footnotesize,labelfont=footnotesize]{caption}
\usepackage[left=1.62cm,right=1.62cm,top=1.9 cm]{geometry}
\def\BibTeX{{\rm B\kern-.05em{\sc i\kern-.025em b}\kern-.08em T\kern-.1667em\lower.7ex\hbox{E}\kern-.125emX}}
\floatname{algorithm}{Algorithm}

\makeatletter
\def\endthebibliography{%
	\def\@noitemerr{\@latex@warning{Empty `thebibliography' environment}}%
	\endlist
}
\makeatother
\usepackage{acronym,booktabs}
\graphicspath{{Pictures/}}

\def\BibTeX{{\rm B\kern-.05em{\sc i\kern-.025em b}\kern-.08em
		T\kern-.1667em\lower.7ex\hbox{E}\kern-.125emX}}
\DeclarePairedDelimiterX\MeijerM[3]{\lparen}{\rparen}%
{\begin{smallmatrix}#1 \\ #2\end{smallmatrix}\delimsize\vert\,#3}

\IEEEoverridecommandlockouts

\normalsize

\ifCLASSINFOpdf
 
\else

\fi

\IEEEoverridecommandlockouts

\begin{document}

\title{\LARGE{RetinaSC: Retinal Semantic Communication Framework for Diabetic Retinopathy Detection}
 \vspace{-0.6em}}
\author{\small{Avani Shinod}\IEEEauthorrefmark{1},~\small{Rangineni Rushitha}\IEEEauthorrefmark{1},~
\small{Asqeem Nas}\IEEEauthorrefmark{1},~
\small{Anirban Bhowal}\IEEEauthorrefmark{1},~
\small{Shikha Baghel}\IEEEauthorrefmark{2}

\\\IEEEauthorblockA{\IEEEauthorrefmark{1}
\small{Department of Electronics and Communication Engineering, NIT Rourkela, Odisha, India-769008.}\\
\IEEEauthorrefmark{2}{Department of Electrical Engineering, IIT Jammu, Jammu, India-181221.}\\
E-mail: \{225EC5024, 225EC5027, 225EC5025, bhowala\}@nitrkl.ac.in, shikha.baghel@iitjammu.ac.in. \vspace{-3em}}
      
\thanks{This work was supported through multiple grants provided by TTDF-DoT under Project Number TTDF/6G/255 and ANRF-ECRG under Project Number ANRF/ECRG/2025/002278/EN.}
}% <-this % stops a space

% 

% The paper headers

\maketitle
% \newline

\begin{abstract}
This paper presents a lightweight semantic communication framework for diabetic retinopathy (DR) grading from fundus images pertaining specifically to the APTOS2019 dataset. The framework includes semantic encoder designed using ResNet34, U-Net, MobileNetV2, and EfficientNet-B0  architectures. Further, the trained models are used to train the same framework for IDRiD dataset using transfer learning approach. Such a framework will enable the telemedicine sector in the early detection of DR and prevent total blindness. The results reveal that for APTOS, task-oriented semantic systems can achieve satisfactory robust performance (more than 0.8 weighted F1 score for ResNet34 and EfficientNet-B0) across different channels. In particular, considering the trade-off between complexity, latency, throughput, and F1 score, MobileNetV2 is the best model with weighted F1 score of 0.76-0.78 for both task and reconstruction-oriented modules and end-to-end latency of only 8.162 ms.

\end{abstract}

% \vspace{-0.1in}
\vspace{-.25cm}

\section{Introduction}
\label{sec:introduction}
 
With the evolution of wireless communication to 6G and Beyond, the focus has shifted from bit-centric to meaning-centric communication \cite{niu2022paradigm}. This paradigm shift has led to the development of semantic communication, which extracts task specific features using deep learning (DL) algorithms to capture the intended meaning and convey these features instead of the entire bit stream, thereby saving bandwidth and other resources \cite{SemComSurvey2023, semanticFeatureExtractDL}. This is more relevant from the perspective of emerging data-centric and application-centric communication use cases, where the accurate delivery of task-relevant information is more important than the faithful reconstruction of the transmitted bits. Semantic communication systems can be either task or reconstruction-oriented \cite{INDISCON}.
 
Healthcare is a critical sector where semantic communication can be widely deployed \cite{healthcare_6g_survey}. Specifically, in rural areas with limited network connectivity, the lack of specialist physicians can lead to improper diagnosis and subsequent casualties. In such scenarios, images or speech signals generated after performing medical tests can be semantically compressed and transmitted through conventional wireless media to specialists in urban areas, where the semantic features can be used for performing the final tasks and subsequent diagnosis. In particular, medical images are large and bandwidth intensive. Hence, semantic communication can be deployed in transmitting only relevant features of the images and utilize them for tasks rather than pixel level reconstruction. One such use case can be the detection of diabetic retinopathy (DR) which is the leading cause of preventable blindness among working adults worldwide, arising as a micro-vascular complication of long term diabetes \cite{DR2025}. Since DR is asymptomatic in its early, most treatable stages, regular retinal screening is essential for timely intervention, yet such screening programs are limited in rural areas.
 
Prior works in image-based semantic communication have focused on domain knowledge-driven models, masked autoencoder-based models, multi-modal smart switching schemes \cite{ChengSurvey_2026}, and diffusion-based models particularly for medical images \cite{diffusionMedicalSemantic}. From the perspective of retinal images in non-semantic framework, research pertains to SegNet-based retinal vessel segmentation \cite{segnetVesselRetinal}, transfer learning-based Swin Transformer encoders for vessel segmentation \cite{swinVesselTransformer}, multi-modal models for DR grading \cite{Messica2026}, and deep learning-based DR grading \cite{Retina2024}. However, investigation of retinal image-based semantic communication framework in both task-oriented or reconstruction oriented modes is lacking. Moreover, the aforementioned architectures are not lightweight and are expensive to deploy on practical edge devices. Further, the use of transfer learning approach for adaptation of the model to a small and imbalanced clinical dataset is also lacking. Such validations are necessary for practical deployment of retinal image-based semantic frameworks in healthcare domain.

The research gaps motivate us to propose a semantic framework in which lightweight architectures such as MobileNetV2 \cite{mobilenetv2_2018}, EfficientNet-B0 \cite{efficientnet2019}, ResNet34 \cite{resnet2016}, and U-Net \cite{UNet} can be leveraged to extract the semantic features. Specifically, MobileNetV2, EfficientNet-B0, and ResNet34 are feasible for medical image classification tasks; while the encoder decoder design of U-Net makes it highly effective for dense pixel-level tasks, including both semantic segmentation and image reconstruction. The feasibility of deployment of such architectures in different scenarios has been investigated in this paper. For DR grading and macular edema (ME) classification, Indian Diabetic Retinopathy Image Dataset (IDRiD) dataset\cite{IDRID2018} is available. However, its limited size and class imbalance motivates us to opt for Asia Pacific Tele-Opthalmology Society (APTOS) 2019 dataset\cite{aptos2019}, and then use pretrained weights to apply transfer learning approach for IDRiD. The key novelty lies in the development of lightweight semantic communication frameworks for DR grading and ME classification, along with an evaluation of practical deployment metrics, including latency and throughput, and a comprehensive ablation study.

Thus, the major contributions can be enumerated as follows: 1) Design of a retinal semantic communication (RetinaSC) framework for detection of DR, working in both task-oriented and reconstruction mode on APTOS2019 and IDRiD datasets; 2) Utilize ResNet34, MobileNetV2, EfficientNet-B0, and U-Net to design the semantic encoders and perform detailed comparison; 3) Performance evaluation over additive white Gaussian noise (AWGN), Rayleigh, and Rician channels in terms of weighted F1 score, latency, throughput, and structural similarity index metric (SSIM) for reconstruction-oriented models; and 4) Detailed ablation study of the framework.

\section{System Model}
\label{sec:system_model}
\begin{figure*}[!t]
\centering
\includegraphics[width=17 cm, height=7 cm]{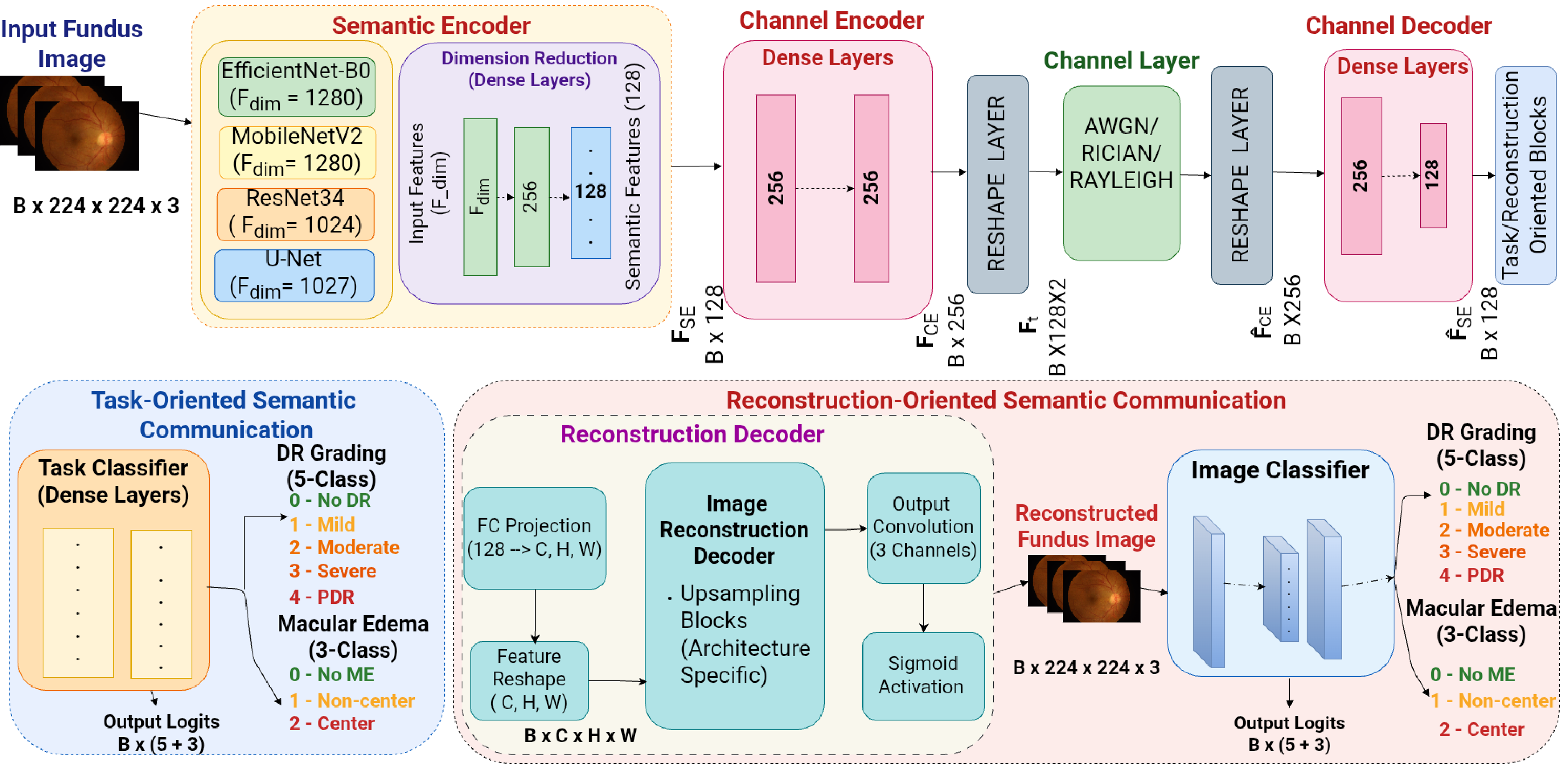}
\caption{RetinaSC semantic framework for task and reconstruction-oriented operations.}
\label{fig:overall_architecture}
\vspace*{-0.15in}
\end{figure*}

The proposed RetinaSC framework, pictorially illustrated in Fig. \ref{fig:overall_architecture}, can operate in either task or reconstruction-oriented mode. Each input fundus image $I \in \mathbb{R}^{p \times q \times n}$ is normalized and processed by the semantic encoder in batches of size $B\times p \times q \times n$, where $p=q=224$ and $n=3$ for both APTOS and IDRiD datasets, corresponding to the RGB color channels.
\subsection{Semantic Transmitter} Semantic transmitter primarily consists of semantic encoder which comprises different architectures for extracting features and a channel encoder to format the features according to the requirements of the transmission channel.
\subsubsection{Semantic Encoder }
The semantic encoder ${\rm T}_{\alpha}^{\rm S}(\cdot)$, parameterized by $\alpha$, can have any one of the four backbone feature extractors (ResNet34, U-Net, MobileNetV2, EfficientNet-B0). Specifically, ResNet34 \cite{resnet2016} consists of 34 weighted layers arranged in four stages of residual blocks, with identity shortcut connections bypassing every pair of convolutional layers. The U-Net encoder \cite{UNet} consists of a contracting path of repeated 3×3 convolution pairs interleaved with 2×2 max-pooling across five stages, that progressively captures hierarchical multi-scale image features. MobileNetV2 \cite{mobilenetv2_2018} utilizes inverted residual blocks with linear bottlenecks and depthwise separable convolutions to achieve computational efficiency while preserving representational capability. EfficientNet-B0 \cite{efficientnet2019} is composed of mobile inverted bottleneck convolution (MBConv) blocks with squeeze and excitation attention, and employs compound scaling to jointly optimize depth, width, and input resolution. The extracted feature representation has dimensions of $\mathbf{F}\in\mathbb{R}^{B\times F_{\rm dim}}$, where $F_{\rm dim}$ depends on the particular architecture. Specifically, $F_{\rm dim}= 1280$ for MobileNetV2 and EfficientNet-B0, whereas $F_{\rm dim}=1027$ and $1024$ for U-Net and ResNet34, respectively. This feature representation is subsequently passed through two dense layers to generate a semantic vector $\mathbf{F}_{\rm SE}\in \mathbb{R}^{B\times 128}$.

\subsubsection{Channel Encoder}
The Channel Encoder $\rm{T}_{\beta}^{\rm C}(\cdot)$, parameterized by $\mathbf{\beta}$, comprises two dense layers each with 256 nodes, and ReLU activation in the first dense layer. This module transforms $\mathbf{F}_{\rm SE} \in \mathbb{R}^{B\times 128}$ to $\mathbf{F}_{\rm CE} \in \mathbb{R}^{B\times 2C_{\rm u}}$, with channel units, $C_{\rm u}=128$. The factor of $2$ signifies in-phase and quadrature phase components of the features.

The output from the channel encoder is passed through a reshape layer to make the feature data compatible for transmission through a wireless channel, resulting in $\mathbf{F}_{\rm t} \in \mathbb{R}^{B\times C_{\rm u}\times 2}$. To ensure consistent transmit power across different channel conditions, the transmitted symbols are normalized.

% \vspace{-0.3cm}
\subsection{Channel Layer}

The channel layer is used to emulate the physical channel. The received signal after passing through the channel layer can be evaluated as
$\mathbf{Y}=\mathbf{H}\mathbf{F}_{\rm t}+\mathbf{N}$, where the entries of $\mathbf{H}$ represents the fading coefficients following Rayleigh or Rician distribution and $\mathbf{N}$ has entries following AWGN. For AWGN channel, only noise will be present. We consider unit average transmit power. 

\subsection{Semantic Receiver} Semantic receiver consists of two sub-modules: channel decoder (${\rm{R}}_{\gamma}^{\rm D}{(\cdot)}$) and task classifier. The received signal $\mathbf{Y}$ from the channel is reshaped to $\hat{\mathbf{F}}_{\rm CE} \in \mathbb{R}^{B\times 2C_{\rm u}}$.

\subsubsection{Channel Decoder}
\label{subsubsec:Ch_dec}

The channel decoder, governed by $\gamma$ parameter, processes the noisy received features $\hat{\mathbf{F}}_{\rm CE}$ and maps them to $\hat{\mathbf{F}}_{\rm SE} \in \mathbb{R}^{B\times 128}$. The channel decoder comprises two dense layers, first one with 256 neurons and RELU activation, and the second one having 128 neurons. The output $\hat{\mathbf{F}}_{\rm SE}$ is passed subsequently through a task-based classifier or reconstruction module.
\subsubsection{Task Classifier}
\label{subsubsec:Em_clssifi}

The task classifier takes $\hat{\mathbf{F}}_{\rm SE}$ as input and utilizes two different classification heads to predict the label for DR grading and ME (only for IDRiD). Each classification head consists of two dense layers that project the recovered semantic features to the corresponding output classes, followed by a Softmax activation to obtain the class probabilities.

\subsubsection{Reconstruction-Oriented Module}

In the reconstruction-oriented framework, the output from the channel decoder $\mathbf{\hat{F}}_{\rm SE}$ is passed to an architecture specific reconstruction decoder $\rm R _{\delta}^{\rm I}(\cdot)$, which reconstructs the input image. Initially $\mathbf{\hat{F}}_{\rm SE}$ is projected through a fully connected layer and reshaped into an initial low-resolution feature map of dimension $B\times C \times H \times W$, where $C, H$, and $W$ are determined by the selected architecture. Subsequently, a sequence of architecture dependent upsampling blocks progressively restores the spatial resolution while refining the feature representation. Depending on the selected architecture, these blocks comprise transpose convolutions, interpolation-based upsampling followed by convolution, or skip-connected decoder stages. A final refinement convolution is applied before a $n$ channel output convolution with a Sigmoid activation produces the reconstructed RGB fundus image. The pixel intensities are confined within the interval [0,1]. The reconstructed image is then forwarded to the image classifier $\rm C_{\phi}(\cdot)$ for DR grading and, in the case of the IDRiD dataset, for ME classification.

\section{Dataset and Training Details}
\label{subsec:datasets}

\subsubsection{Dataset}
With a particular focus on the Indian rural scenario, the IDRiD dataset has been used, which is a publicly available fundus dataset providing both DR grade and ME risk annotations \cite{IDRID2018}. It contains a total of 516 retinal fundus images, collected at an eye clinic in Nanded, Maharashtra, India, and split into 413 images for training and 103 images for testing, with an appropriate mixture of disease stratification maintained across both sets. However, IDRiD has limited training images and an imbalanced class distribution, which motivates us to opt for APTOS dataset. The APTOS 2019 dataset is a lot bigger than IDRiD and contains high resolution fundus images collected from rural people \cite{aptos2019}. The dataset was created for the APTOS 2019 Blindness Detection competition, aimed at automatic screening for DR. It contains retinal fundus (back-of-eye) photographs, consisting of a total of 3662 images, divided into 2930 training images, 366 test images, and 366 validation images. The images were collected by Aravind Eye Hospital in India, largely from patients in rural areas with limited access to eye-care specialists and reflects real-world screening conditions (varying camera types, lighting, and image quality).

\subsection{Training Procedure (APTOS)}
\label{subsec:train_combined}

Algorithm~\ref{algo:combined_training} illustrates the training process of both task-oriented and reconstruction-oriented systems using APTOS dataset. A flag is used to indicate task or reconstruction-oriented mode, enabling the system to determine the appropriate processing strategy for the information.

\begin{algorithm}[t]
\caption{Training Procedure (APTOS): Task-Oriented and Reconstruction-Based Systems}
\label{algo:combined_training}
\begin{algorithmic}[1]
\STATE \textbf{Initialization:} Initialize parameters $\Theta^{(0)}$, $i=0$.
\STATE \textbf{Input:} Fundus images $I$ with labels $y_{\text{grade}}$ from train set, fading channel $\mathbf{H}$ and noise $\mathbf{N}$, system type $\in \{\text{task-oriented}, \text{reconstruction-based}\}$.
\STATE Normalize images $I$ from the train set.
\WHILE{Training loss is not converged}
    \STATE Encode using semantic encoder followed by channel encoder to generate features: $\mathbf{F}_{\rm CE} = \mathrm{T}^{\rm C}_{\beta}\left(\mathrm{T}^{\rm S}_{\alpha}(I)\right)$.
    \STATE Reshape $\mathbf{F}_{\rm CE}$ to obtain $\mathbf{F}_{\rm t}$ and transmit it over Rayleigh, Rician, or AWGN channel to receive $\mathbf{Y}$.
    \STATE Reshape $\mathbf{Y}$ to obtain $\hat{\mathbf{F}}_{\rm CE}$ and further decode received semantic features as $\hat{\mathbf{F}}_{\rm SE} = \mathrm{R}^{\rm D}_{\gamma}(\hat{\mathbf{F}}_{\rm CE})$.
    \IF{system type is task-oriented}
        \STATE Compute DR grade logits directly from $\hat{\mathbf{F}}_{\rm SE}$.
        \STATE Compute total loss (where $\mathcal{L}_{\text{CE}}$ and $\mathcal{L}_{\text{sem}}$ denote the cross-entropy classification loss and semantic feature consistency loss, respectively, $\lambda$ is the weight factor):
        \begin{equation}
            \mathcal{L}(\Theta) = \mathcal{L}_{\text{CE}}(y_{\text{grade}}, \hat{y}_{\text{grade}}) + \lambda \, \mathcal{L}_{\text{sem}}\left(\mathrm{T}^{\rm S}_{\alpha}(I), \hat{\mathbf{F}}_{\rm SE}\right).
            \label{eq:loss_task}
        \end{equation}
    \ELSE
        \STATE Reconstruct the image $\hat{I} = \mathrm{R}^{\rm I}_{\delta}(\hat{\mathbf{F}}_{\rm SE})$ and compute DR grade logits $\hat{y}_{\rm grade}$ from $\mathrm{C}_{\phi}(\hat{I})$.
        \STATE Compute total loss (where $\mathcal{L}_{\text{recon}}$ denotes the reconstruction loss and $\lambda_{\rm r}$ is the reconstruction loss weighting coefficient):
        \begin{equation}
            \mathcal{L}(\Theta) = \mathcal{L}_{\text{CE}}(y_{\text{grade}}, \hat{y}_{\text{grade}}) + \lambda_{\rm r} \, \mathcal{L}_{\text{recon}}(I, \hat{I}).
            \label{eq:loss_recon}
        \end{equation}
    \ENDIF
    \STATE Update model parameters ($\eta$ and $\nabla$ representing the learning rate and nabla operator, respectively):
    \begin{equation}
        \Theta^{(i+1)} \leftarrow \Theta^{(i)} - \eta \nabla_{\Theta^{(i)}} \mathcal{L}(\Theta).
        \label{eq:update_combined}
    \end{equation}
\ENDWHILE
\STATE \textbf{Output:} Trained networks $\mathrm{T}^{\rm C}_{\beta}(\cdot)$, $\mathrm{T}^{\rm S}_{\alpha}(\cdot)$, $\mathrm{R}^{\rm D}_{\gamma}(\cdot)$, and either the classification head (task-oriented) or $\mathrm{R}^{\rm I}_{\delta}(\cdot)$ and $\mathrm{C}_{\phi}(\cdot)$ (reconstruction-based).
\end{algorithmic}
\end{algorithm}

\subsection{Transfer Learning to IDRiD}
\label{subsec:transfer_learning}

Due to the aforementioned limitations of IDRiD, both task-oriented and reconstruction-based systems are adjusted from their APTOS-trained checkpoints to work with IDRiD using transfer learning approach. The parts of the model carried over from APTOS are fine tuned slowly and carefully, while the new edema branch is trained swiftly.

\begin{algorithm}[t]
\caption{Transfer Learning Procedure (APTOS $\rightarrow$ IDRiD)}
\label{algo:transfer}
\begin{algorithmic}[1]
\STATE \textbf{Initialization:} Load pretrained APTOS weights into $\mathrm{T}^{\rm S}_{\alpha}(\cdot)$, $\mathrm{T}^{\rm C}_{\beta}(\cdot)$, $\mathrm{R}^{\rm D}_{\gamma}(\cdot)$, and the DR grading branch of the classifier (or $\mathrm{R}^{\rm I}_{\delta}(\cdot)$ and $\mathrm{C}_{\phi}(\cdot)$ for the reconstruction-based system); re-initialize the edema branch; $i=0$.
\STATE \textbf{Input:} IDRiD images $I$ with labels $y_{\text{grade}}, y_{\text{edema}}$ from train set, learning rate $\eta_{\text{ft}}$, edema-branch learning rate $\eta_{\text{head}}$.
\STATE Construct a class-balanced sampler over the IDRiD train set.
\WHILE{Training loss is not converged}
    \STATE Perform a forward pass through the initialized semantic communication pipeline to obtain reconstructed semantic features (task-oriented) or reconstructed images (reconstruction-oriented), followed by DR and edema predictions to obtain $\hat{y}_{\text{grade}}$ and $\hat{y}_{\text{edema}}$. 
    \STATE Compute total loss $\mathcal{L}(\Theta)$ as in~\eqref{eq:loss_task} or~\eqref{eq:loss_recon}, with both DR and edema terms active.
    \STATE Update pretrained model parameters $\Theta_{\text{DR}}$ at rate $\eta_{\text{ft}}$ and edema-branch parameters $\Theta_{\text{edema}}$ at rate $\eta_{\text{head}}$:
    \begin{equation}
    \begin{split}
    \Theta^{(i+1)}_{\text{DR}}
    &\leftarrow \Theta^{(i)}_{\text{DR}}
    - \eta_{\text{ft}} \nabla \mathcal{L}(\Theta), \\
    \Theta^{(i+1)}_{\text{edema}}
    &\leftarrow \Theta^{(i)}_{\text{edema}}
    - \eta_{\text{head}} \nabla \mathcal{L}(\Theta).
     \end{split}
    \label{eq:update_transfer}
    \end{equation}
\ENDWHILE
\STATE \textbf{Output:} Fine-tuned networks adapted to IDRiD, including the newly trained edema branch.
\end{algorithmic}
\end{algorithm}

\section{Results and Discussion}

The proposed semantic communication framework is trained over 200 epochs at a fixed signal-to-noise ratio (SNR) of 15 dB, employing early stopping with a patience counter of 30. To validate the robustness of the framework, the testing is conducted over a wide range of SNR varying from -20 dB to 30 dB. For model optimization, two hybrid objective functions comprising mean squared error (MSE) loss and cross entropy (CE) loss are being used. The MSE loss is used for the purpose of image reconstruction and CE Loss for disease classification. To evaluate the retention of structural details in the reconstructed retinal images, the reconstruction performance is measured using SSIM at various SNR levels. Classification performance is assessed using the confusion matrix, specifically weighted F1 score, to offer a thorough evaluation of the ability to predict disease. The training and validation loss curves for task and reconstruction-oriented models converge satisfactorily within 200 epochs. However, they are skipped in this paper due to space constraints.

\subsubsection{Performance of Task and Reconstruction-Oriented Modules with APTOS}

\begin{figure*}[t]
\centering
\begin{subfigure}[b]{0.33\textwidth}
    \centering
    \includegraphics[width=2.3in, height=2in]{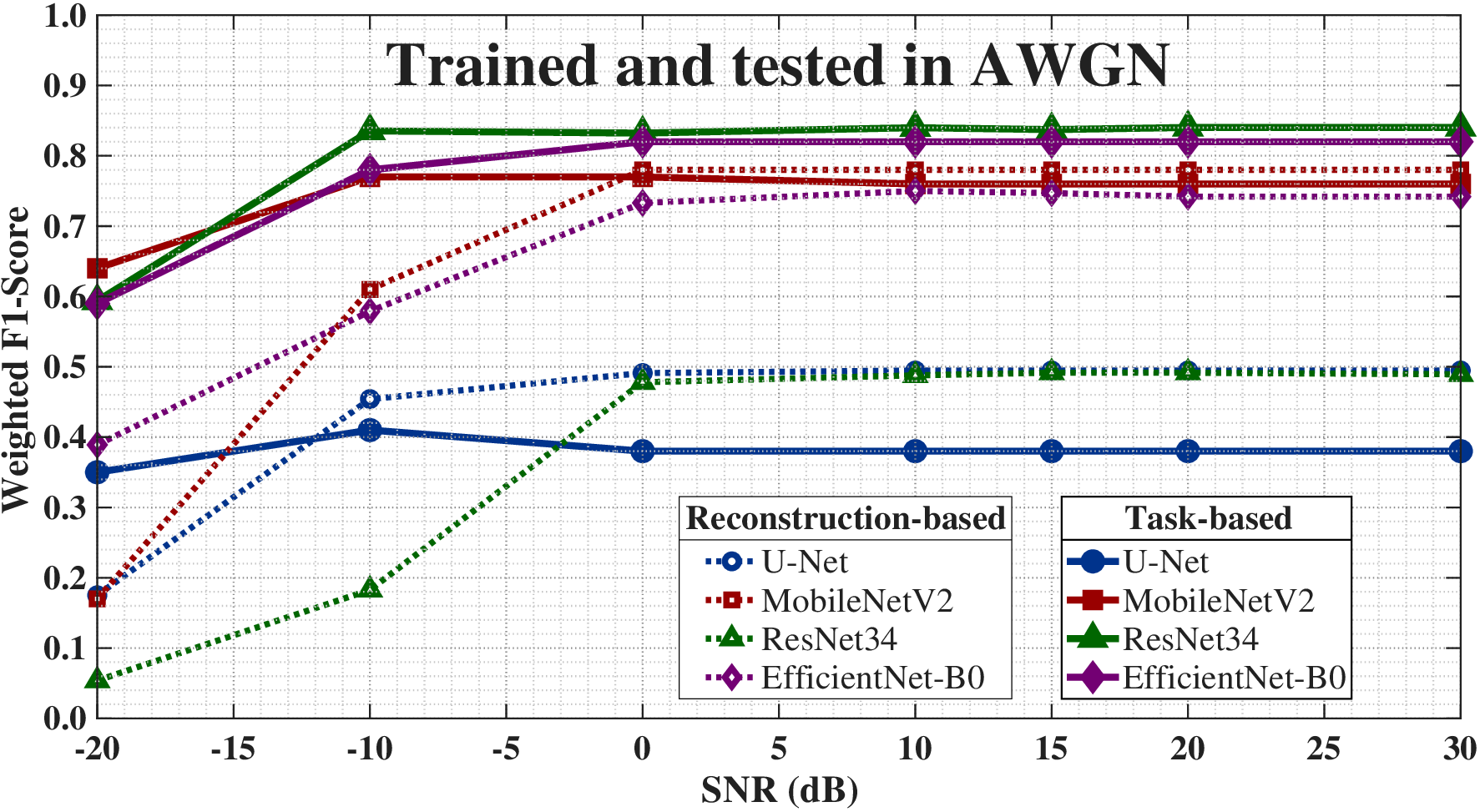}
    \caption{}
    \label{fig:f1_awgn}
\end{subfigure}
\hfill
\begin{subfigure}[b]{0.33\textwidth}
    \centering
    \includegraphics[width=2.3 in, height=2in]{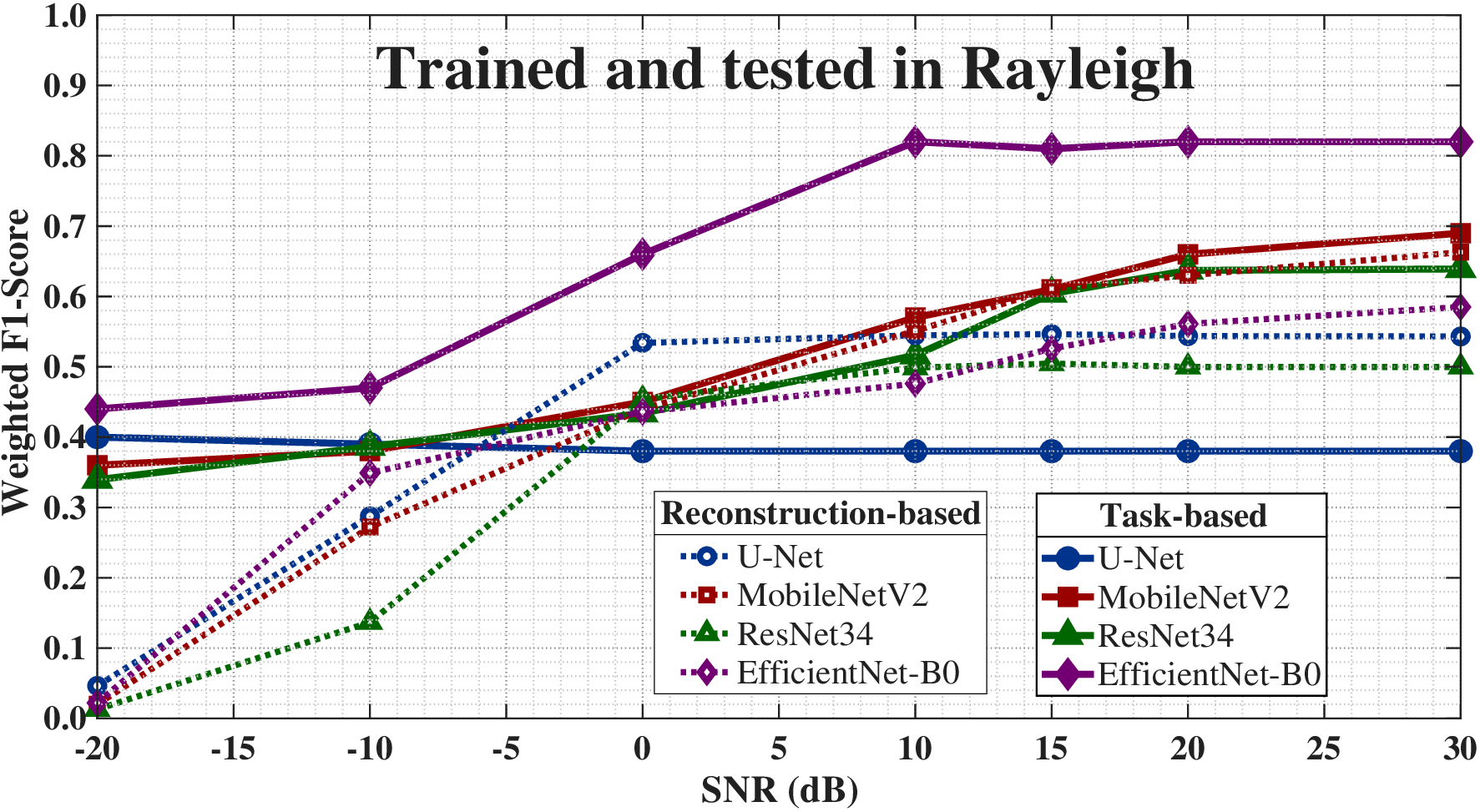}
    \caption{}
    \label{fig:f1_rayleigh}
\end{subfigure}
\hfill
\begin{subfigure}[b]{0.32\textwidth}
    \centering
    \includegraphics[width=2.3in, height=2in]{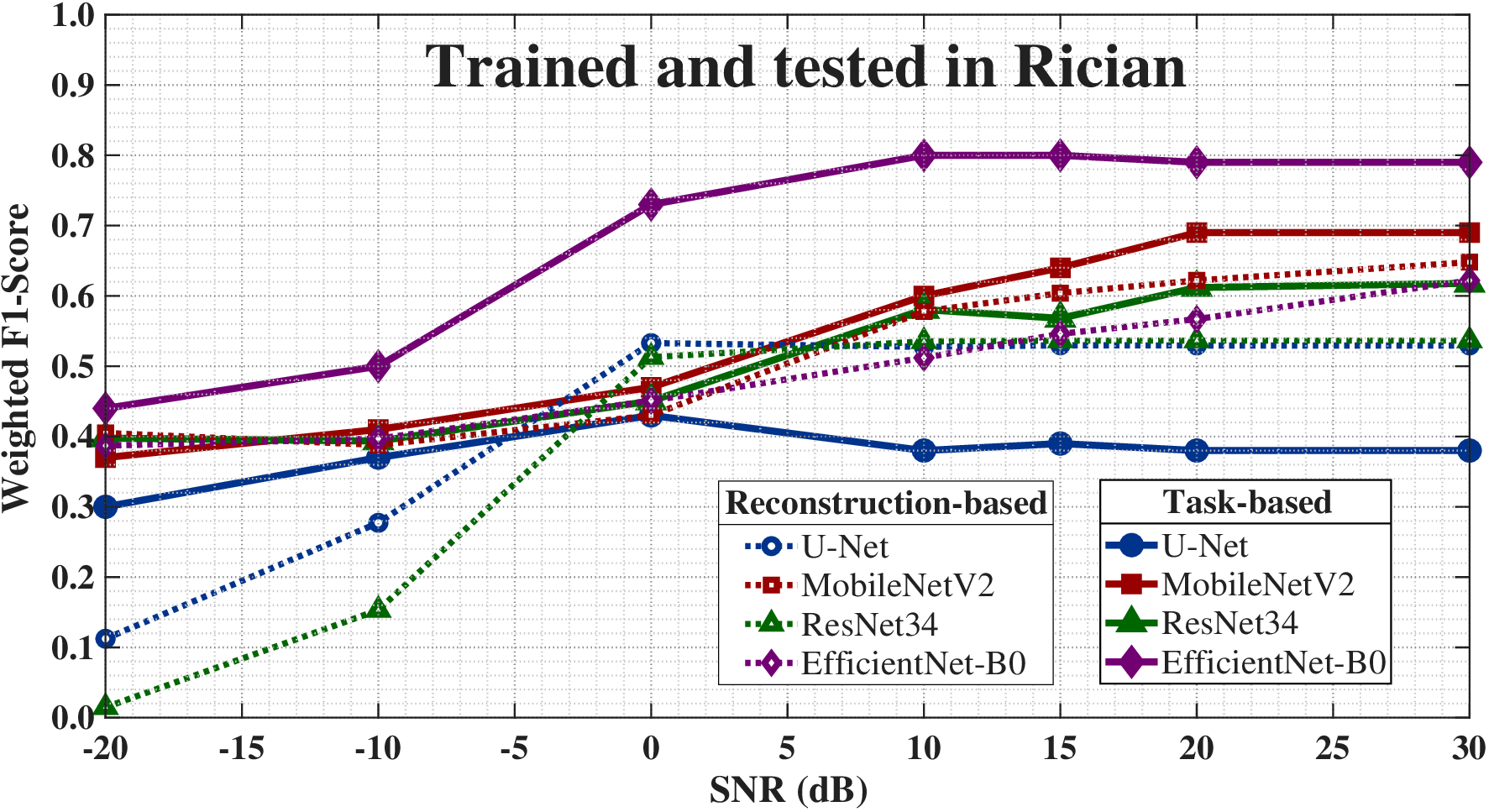}
    \caption{}
    \label{fig:f1_rician}
\end{subfigure}
\caption{Performance evaluation of task and reconstruction-based modes for different architectures in (a) AWGN, (b) Rayleigh, and (c) Rician channel conditions.}
\vspace*{-0.1in}
\label{fig:f1_all_channels}
\end{figure*}

The performance of the four different architectures for both task-oriented and reconstruction-oriented systems is tested across AWGN, Rayleigh, and Rician channels (cf. Fig. \ref{fig:f1_awgn}, \ref{fig:f1_rayleigh}, and \ref{fig:f1_rician}, respectively), after training over the corresponding channels. The performance over AWGN channel is the best with ResNet34 and EfficientNet-B0 achieving peak F1 scores of 0.84 and 0.82, respectively for task-oriented systems. Meanwhile for the same models, the peak F1 scores for ResNet34 and EfficientNet-B0 are 0.64 and 0.82, respectively, for Rayleigh channel, and 0.62 and 0.8, respectively, for Rician channel. Although ResNet34 achieves the highest performance under the AWGN channel, its performance degrades significantly under Rayleigh and Rician fading as the extracted semantic features are more susceptible to multipath-induced distortions. In contrast, EfficientNet-B0 employs compound scaling strategy and efficient squeeze and excitation attention modules to learn compact and discriminative semantic representations, ensuring robustness to fading and consistent classification performance across diverse channel conditions. It is noteworthy that MobileNetV2 can attain peak F1 scores of 0.76, 0.69, and 0.69 over AWGN, Rayleigh, and Rician channels, respectively. This indicates that this architecture prioritizes computational efficiency over representational capacity, leading to reduced discriminative power and increased sensitivity to channel impairments.

However, the reconstruction-oriented models provide contrasting results. MobileNetV2 achieves the highest reconstruction-oriented performance (peak F1 score of 0.78) as its lightweight feature representation is more effectively reversed by the reconstruction decoder, preserving sufficient diagnostic information for classification. EfficientNet-B0 (peak F1 score of 0.75) exhibits slightly lower performance as its semantic features are optimized for discriminative classification rather than image reconstruction. In contrast, ResNet34 with peak F1 score of 0.49 learns highly abstract features that are difficult to reconstruct faithfully, while U-Net with peak F1 score of 0.49 primarily optimizes pixel-level similarity and performs poorly for both task and reconstruction-oriented systems. The trends are similar across all channels.

\captionsetup[subfigure]{labelformat=simple, labelsep=none}
\renewcommand{\thesubfigure}{(\alph{subfigure})}

Since AWGN trained models perform the best, for further analysis, AWGN trained models are considered. Testing is performed over Rayleigh and Rician channels on trained AWGN task-oriented models in Fig. \ref{fig:f1_task_cross} and reconstruction-oriented models in Fig. \ref{fig:f1_reconstruct_cross} to validate the robustness of the proposed framework. EfficientNet-B0 generalizes well with peak F1 scores reaching 0.82 while testing on both Rayleigh and Rician, whereas ResNet34 attains peak F1 scores of 0.84 while testing on other channels, which matches the peak performance of testing on AWGN for task-oriented framework. However, for reconstruction-oriented framework, MobNetV2 generalizes well with peak F1 score of 0.78 in all the three channels, followed closely by EfficientNet-B0 with corresponding score of 0.75 over all the channels. These results prove that the models which performed the best when trained and tested over AWGN, also perform satisfactorily during testing over other channels.

\subsubsection{Transfer Learning on IDRiD Dataset}
\begin{figure*}[t]
\centering

\begin{minipage}[t]{0.31\textwidth}
\centering
\includegraphics[width=\linewidth,height=4.2cm]{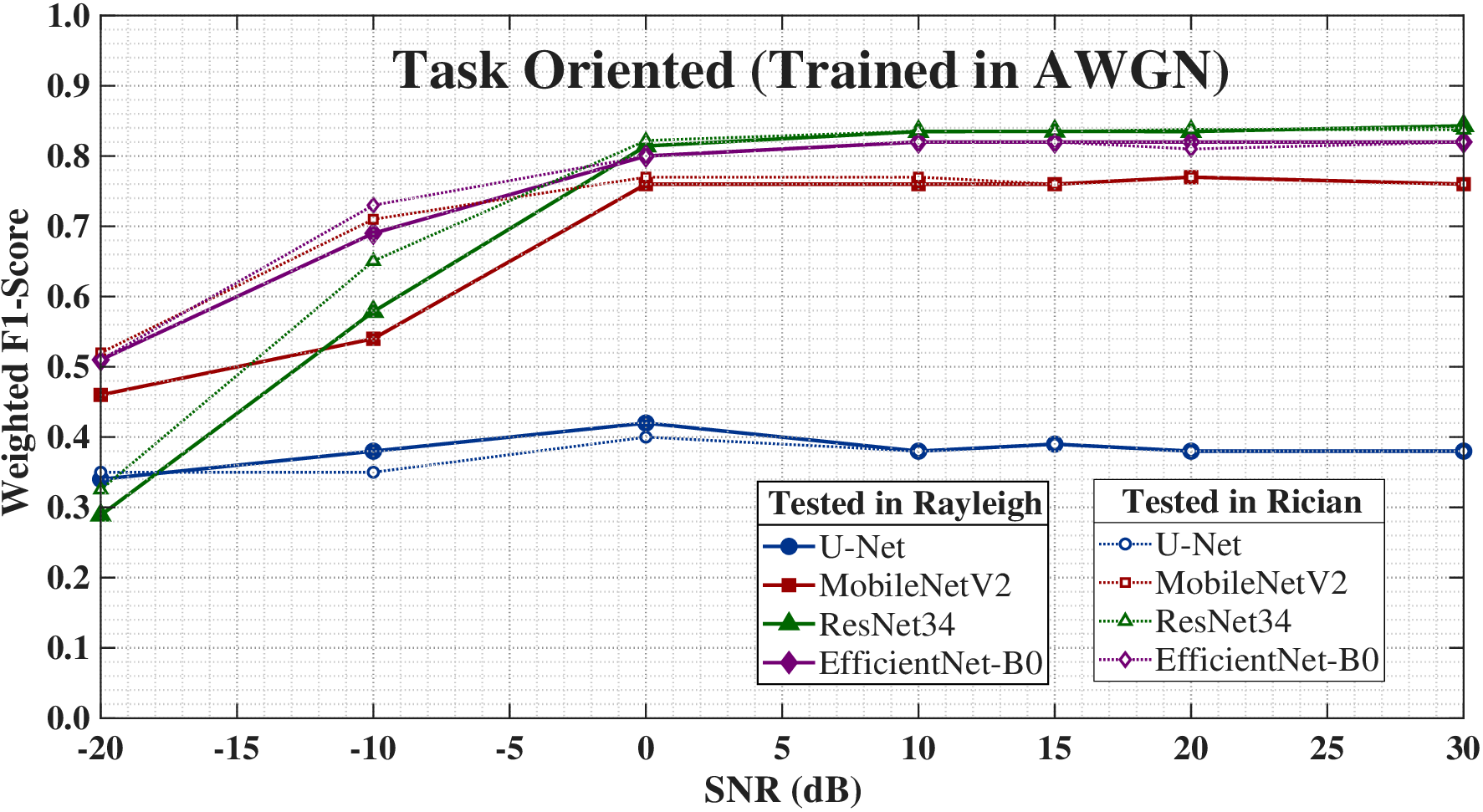}
\captionof{figure}{Performance evaluation of task-oriented models on Rayleigh and Rician channels.}
\label{fig:f1_task_cross}
\end{minipage}
\hfill
\begin{minipage}[t]{0.31\textwidth}
\centering
\includegraphics[width=\linewidth,height=4.2cm]{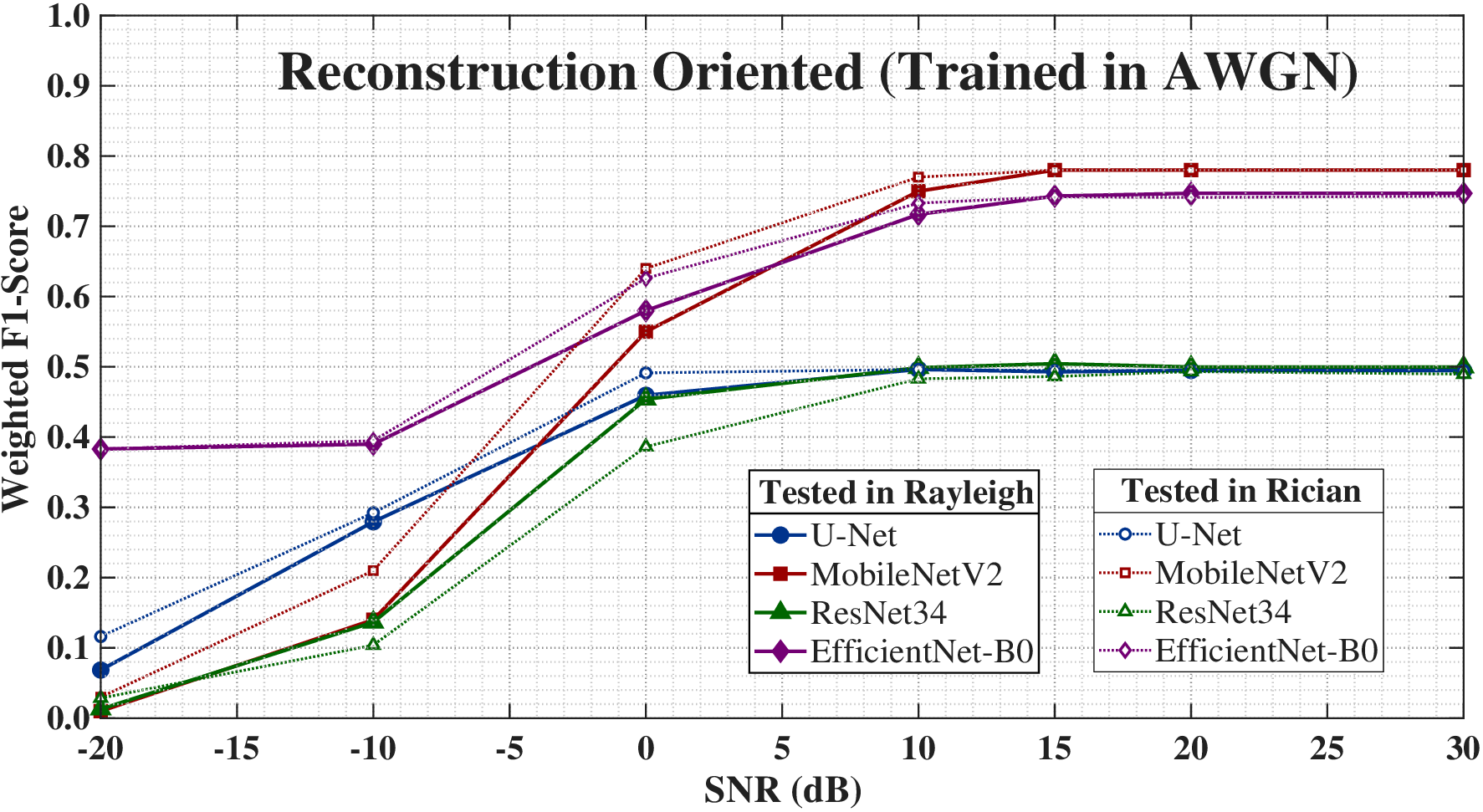}
\captionof{figure}{Performance evaluation of reconstruction-oriented models on Rayleigh and Rician channels.}
\label{fig:f1_reconstruct_cross}
\end{minipage}
\hfill
\begin{minipage}[t]{0.31\textwidth}
\centering
\includegraphics[width=\linewidth,height=4.2cm]{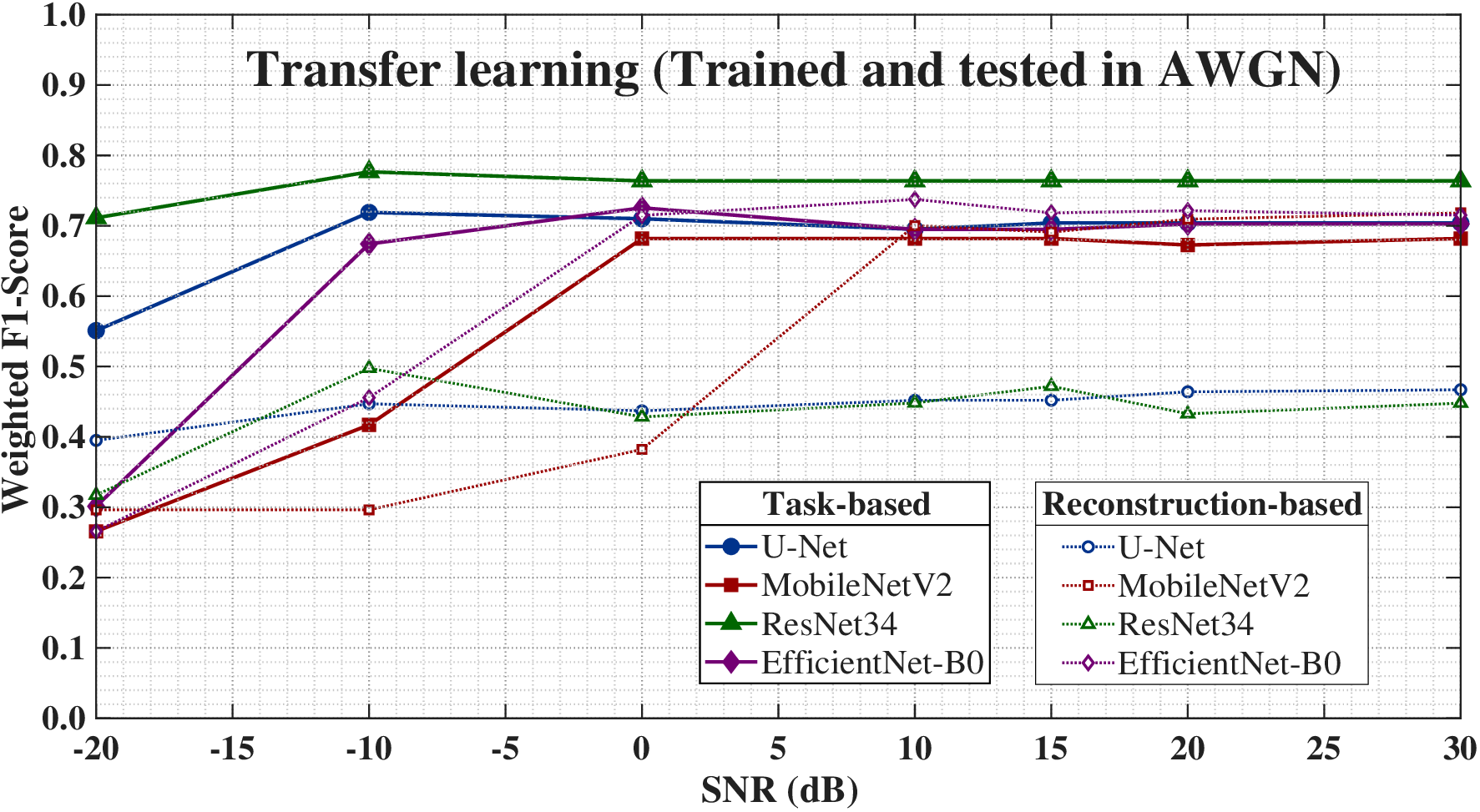}
\captionof{figure}{Transfer learning performance on the IDRiD dataset.}
\label{fig:transferlearning}
\end{minipage}
    \vspace*{-0.15in}
\end{figure*}

Fig. \ref{fig:transferlearning} evaluates a transfer learning scenario in which each of the four architectures, originally trained for the DR classification task, is fine-tuned and evaluated on a different task metric, ME classification, using the IDRiD dataset. Both the task-based and reconstruction-based encoders are trained and tested in AWGN. As the SNR increases beyond 0 dB, most models achieve stable performance, with ResNet34-based task-oriented semantic communication providing the highest ME weighted F1-score (0.76), followed by EfficientNet-B0, MobileNetV2, and U-Net. Compared with the APTOS results, the overall performance on IDRiD is noticeably lower (0.4-0.51 peak weighted F1 score), particularly for DR grading, owing to the smaller dataset size, higher class imbalance, and increased lesion variability in IDRiD. The DR results have not been plotted for IDRiD due to space constraints. Nevertheless, transfer learning significantly improves the performance of all backbones, with U-Net exhibiting a notable gain over its poor DR classification results (0.43 compared to 0.2 peak weighted F1 scores with training over IDRiD), indicating that the proposed framework also generalizes well in related retinal disease analysis tasks.

\subsubsection{Comparison with Baseline Models}
Table \ref{tab:baseline_compare} tabulates end-to-end latency, throughput, SSIM, and weighted F1 scores for all four semantic architectures for both the task-oriented and reconstruction-oriented configurations. Apart from these, benchmark systems have also been included for comparison, which directly perform DR grading on the APTOS dataset without semantic transmission or channel impairments. Task-oriented models are consistently faster than their reconstruction-oriented counterparts, with latency roughly 1.5-2.5 times higher for reconstruction across every model. Specifically, the end-to-end latency of task-oriented models are similar as compared to non-semantic benchmarks, with marginal drop or improvement in peak weighted F1 scores, except U-Net model. U-Net model is particularly suitable for pixel-level reconstruction, which is justified by the highest SSIM value among all reconstruction-oriented models. Thus, it can be inferred that MobNetV2 offers the best trade-off between computational efficiency and performance, exhibiting the lowest latency, highest throughput, and robust classification performance for both task-oriented and reconstruction-oriented semantic communication frameworks.

\begin{table}[]
\centering
\caption{Comparison of the proposed models with baseline models.}
\label{tab:baseline_compare}
\begin{tabular}{|l|l|l|l|l|}
\hline
\textbf{Model} & \textbf{End-to-End} & \textbf{End-to-End} & \textbf{SSIM} &\textbf{Peak}\\
 & \textbf{Latency} & \textbf{Throughput } &  &\textbf{Weighted} \\
 &\textbf{(ms)} &(\textbf{img/s})& &\textbf{F1 Score}\\
\hline

\multicolumn{5}{|c|}{\textbf{Task-Oriented Semantic Communication}} \\ \hline
EfficientNet-B0 & 10.191 & 784.99 & - & 0.82 \\ \hline
MobileNetV2  & 8.162  & 980.16 & - & 0.76 \\ \hline
ResNet34       & 17.915 & 446.60 & - & 0.84 \\ \hline
U-Net         & 16.061 & 498.10 & - & 0.38 \\ \hline

\multicolumn{5}{|c|}{\textbf{Reconstruction-Oriented Semantic Communication}} \\ \hline
EfficientNet-B0 & 17.021 & 467.18 & 0.377 & 0.75 \\ \hline
MobileNetV2  & 14.738 & 539.89 & 0.429 & 0.78 \\ \hline
ResNet34       & 41.387 & 193.30 & 0.578 & 0.49 \\ \hline
U-Net         & 41.708 & 191.80 & 0.806 & 0.49 \\ \hline

\multicolumn{5}{|c|}{\textbf{Non-Semantic Baseline Models}} \\ \hline
EfficientNet-B0 & 9.332 & 860.94 & - & 0.80 \\ \hline
MobileNetV2  & 7.471 & 1070.75 & - &  0.77\\ \hline
ResNet34       & 17.809 & 449.2 & - & 0.798\\ \hline
U-Net         & 15.388 & 519.9 & - & 0.711\\ \hline

\end{tabular}
\end{table}

\subsubsection{Ablation Study}
An ablation study is presented in Table \ref{tab:my-table-2}, where heavy, normal, and light compression indicates the semantic latent count as 64, 128, and 256, respectively. No expansion signifies the latent count as 128 for channel encoder output. Deep classifier represents two levels of dense layers with the feature count being reduced from 128 to 64, and then to 5. For all the task-oriented models, the semantic features remain discriminative under varying levels of compression, demonstrating the robustness of the learned latent representations. A deeper level of classifier is also unnecessary as evident from the results. However, for reconstruction-oriented models, heavy compression is particularly beneficial for EfficientNet-B0 and MobileNetV2. Thus, we can infer that compact latent representations simplify the reconstruction process for such lightweight architectures. Performance remains insensitive to channel encoder complexity.

\begin{table}[]
\centering
\caption{Ablation analysis of the different models.}
\label{tab:my-table-2}
\begin{tabular}{|l|>{\centering\arraybackslash}p{1cm}|>{\centering\arraybackslash}p{1cm}|>
{\centering\arraybackslash}p{1cm}|>
{\centering\arraybackslash}p{0.8cm}|>{\centering\arraybackslash}p{0.8cm}|}
\hline
\textbf{Model} & \multicolumn{5}{c|}{\textbf{Peak Weighted F1 Score}} \\ \cline{2-6}
 & \textbf{Heavy Compression} &\textbf{Normal Compression} & \textbf{Light Compression} & \textbf{No Expansion} & \textbf{Deep Classifier} \\ \hline
\multicolumn{6}{|c|}{\textbf{Task-Oriented Semantic Communication}} \\ \hline
EfficientNet-B0 & 0.83 &0.82 &0.84 & 0.82 & 0.83 \\ \hline
MobileNetV2  & 0.78 &0.76 &0.78 & 0.81 & 0.77 \\ \hline
ResNet34       & 0.79 &0.83 &0.80 & 0.80 & 0.81 \\ \hline
U-Net        & 0.54 &0.38 &0.54 & 0.54 & 0.54 \\ \hline
\multicolumn{6}{|c|}{\textbf{Reconstruction-Oriented Semantic Communication}} \\ \hline
EfficientNet-B0 & 0.81 &0.75 &0.75 & 0.82 & 0.79 \\ \hline
MobileNetV2  & 0.83 &0.78 &0.81 & 0.82 & 0.81 \\ \hline
ResNet34       & 0.55 &0.49 &0.53 & 0.52 & 0.50 \\ \hline
U-Net        & 0.55 &0.49 &0.55 & 0.55 & 0.55 \\ \hline
\end{tabular}
\end{table}

\section{Conclusion}
\label{sec:conclusion}

This paper compared a task-oriented and a reconstruction-based semantic communication framework for DR and ME grading, by leveraging four different semantic encoder architectures. The performance evaluation was done in terms of weighted F1 score over AWGN, Rayleigh, and Rician channels using APTOS and IDRiD datasets. It was revealed that MobileNetV2 offers the best trade-off
between computational efficiency and performance, exhibiting
the lowest end-to-end latency (8.162 ms), highest throughput, and robust classification performance (0.76-0.78 weighted F1 score) across both task-oriented and reconstruction-oriented semantic communication frameworks. It maintains similar performance as compared to non-semantic baseline, which yields weighted F1 score of 0.77 and end-to-end latency of 7.471 ms.

\bibliographystyle{IEEEtran}
\bibliography{Ref}

\end{document}